# Integrated information and dimensionality in continuous attractor dynamics

Satohiro Tajima[1,2,*], Ryota Kanai[3]

1. University of Geneva, Geneva, Switzerland.
2. JST PRESTO, Saitama, Japan.
3. Araya Brain Imaging, Tokyo, Japan.

*satohiro.tajima@gmail.com

## Abstract

There has been increasing interest in the integrated information theory (IIT) of consciousness, which hypothesizes that consciousness is integrated information within neuronal dynamics. However, the current formulation of IIT poses both practical and theoretical problems when we aim to empirically test the theory by computing integrated information from neuronal signals. For example, measuring integrated information requires observing all the elements in the considered system at the same time, but this is practically rather difficult. In addition, the interpretation of the spatial partition needed to compute integrated information becomes vague in continuous time-series variables due to a general property of nonlinear dynamical systems known as "embedding." Here, we propose that some aspects of such problems are resolved by considering the topological dimensionality of shared attractor dynamics as an indicator of integrated information in continuous attractor dynamics. In this formulation, the effects of unobserved nodes on the attractor dynamics can be reconstructed using a technique called delay embedding, which allows us to identify the dimensionality of an embedded attractor from partial observations. We propose that the topological dimensionality represents a critical property of integrated information, as it is invariant to general coordinate transformations. We illustrate this new framework with simple examples and discuss how it fits together with recent findings based on neural recordings from awake and anesthetized animals. This topological approach extends the existing notions of IIT to continuous dynamical systems and offers a much-needed framework for testing the theory with experimental data by substantially relaxing the conditions required for evaluating integrated information in real neural systems.

## Introduction

There has been increasing interest in the integrated information theory (IIT) of consciousness. The central hypothesis of IIT is that consciousness is integrated information within collective neuronal dynamics [1–5]. An attractive aspect of IIT is that it could relate basic properties of subjective experiences in consciousness to the physical mechanisms of biological (and even artificial) dynamical systems via the information theoretic framework [5]. Among the theories of consciousness, IIT is relatively new and still awaits empirical verification. To examine IIT with empirical neural recordings, however, its current implementation needs to address several issues from both practical and theoretical viewpoints. Although empirical studies have reported neural phenomena for which IIT could provide consistent explanations [6–10], it is still challenging to test the necessity of IIT directly with empirical datasets under its current formulation. For example, measuring integrated information in a rigorous sense requires observing all the elements at the same time, which imposes a serious bottleneck in testing the theory with neural recordings in living organisms. Moreover, as we discuss later in this paper, the interpretation of spatial partitioning becomes unclear when we regard a time sequence of continuous variables as a unit of "state" due to the local observability known as "embedding" [11,12] in general continuous dynamical systems.

Here, we discuss an alternative implementation of IIT that could resolve some aspects of those problems. A key idea in our formulation is to index the integrated information in terms of the topological dimensionality of shared attractor dynamics. In this formulation, the effects of unobserved nodes on the attractor dynamics could be reconstructed using a technique called delay embedding. Remarkably, considering topological properties allows us to make use of, rather than suffer from, the puzzling effects of embedding in continuous dynamical systems. We illustrate how this formulation works with simple examples and discuss its relevance to the original formulation of IIT and our recent empirical findings from awake and anesthetized animals [13].

The aim of the present perspective article is to illustrate the basic idea behind our formulation. For this purpose, we focus on intuitive rather than rigorous mathematical descriptions.

## Topological dimensionality as an indicator of integrated information in continuous dynamical systems

To illustrate our formulation, let us consider simple dynamical systems consisting of only two nodes, with values of $x_1$ and $x_2$ (**Figure 1**). Suppose that each of $x_1$ and $x_2$ has self-feedback, which is generally nonlinear. For the sake of simplicity, here we assume that each node's value is defined in one-dimensional continuous space (e.g., $x_1, x_2 \in \mathbb{R}$), but the subsequent arguments are valid for general cases in which each node value is defined in higher-dimensional spaces.

First, let us begin by considering a mutually interacting system (**Figure 1a-i**). We assume the system dynamics to be described with deterministic difference equations as (similar arguments apply to the cases with ordinary differential equations)

$$x_1^t = f(x_1^{t-1}, x_2^{t-1}), \qquad (1)$$

$$x_2^t = g(x_1^{t-1}, x_2^{t-1}), \qquad (2)$$

where $f$ and $g$ are arbitrary continuous functions and $t$ denotes an arbitrary time point. In a general nonlinear system, the state $(x_1^t, x_2^t)$ could be distributed across an at most 2-dimensional manifold $A$ (the light gray

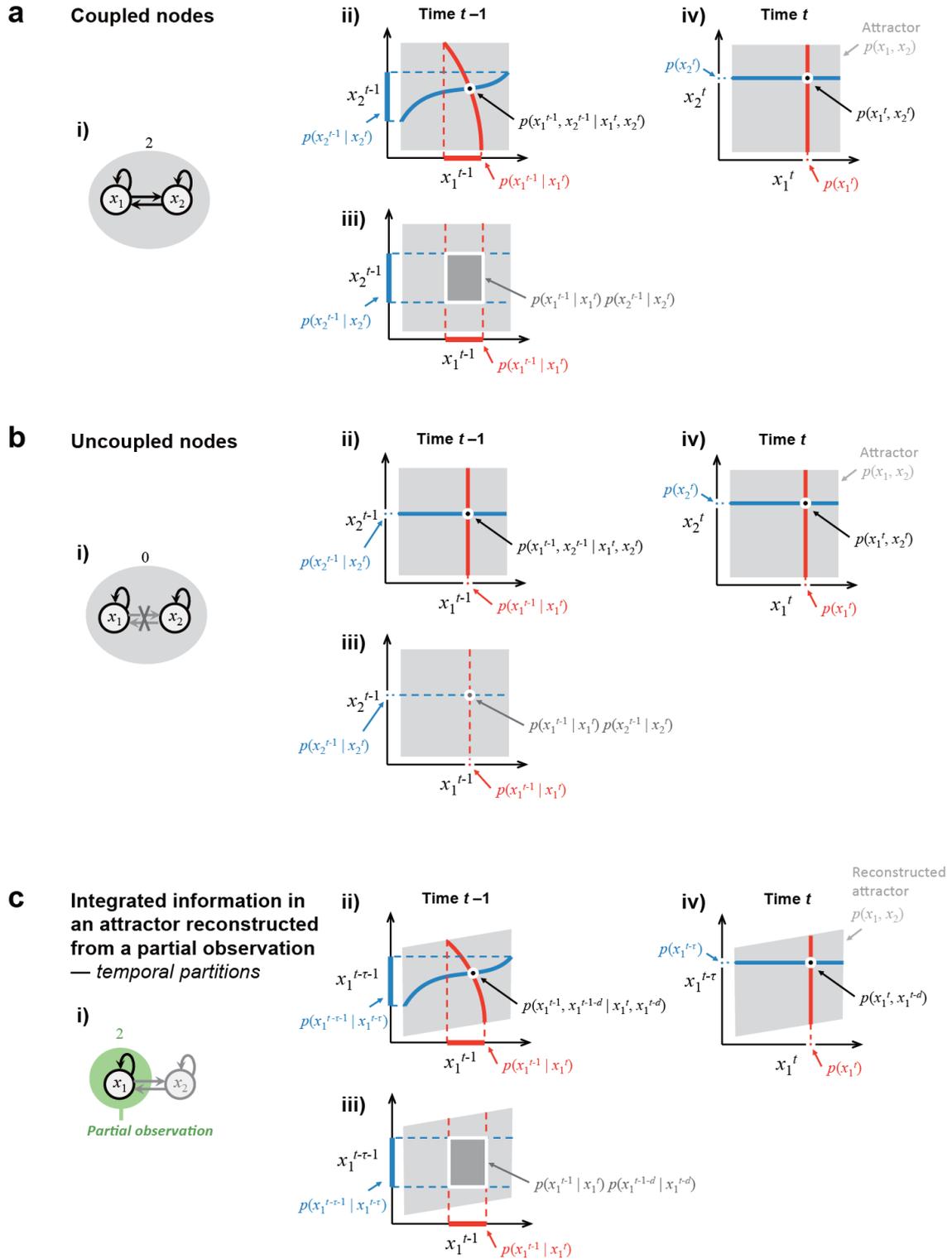

**Figure 1**

Schematic illustrations for the dimensionality-based index of integrated information.
(a) A system with mutually interacting nodes.
(b) A system comprising two disconnected nodes.
(c) A partial observation of the system with mutually interacting nodes (the same system as in panel a).
Insets: i) The schematic of the systems; ii) The inferred past states at time $t-1$; iii) The inferred past state at time $t-1$, based on a partitioned observation; iv) The current states.

square in the figure) in the phase space of $(x_1, x_2)$. For convenience, we call $A$ an "attractor" when the state stays within $A$ for a sufficiently long time. After a sufficient duration away from the initial state, attractor $A$ provides a support of the joint probability density distribution $p(x_1, x_2)$.

Suppose that we could identify both nodes' values $(x_1^t, x_2^t)$ at time $t$ by observing them simultaneously (i.e., we could make the joint probability density distribution $p(x_1^t, x_2^t)$ be a delta function through the observation). If we consider the past state of those nodes, $(x_1^{t-1}, x_2^{t-1})$, based on this observation, the uncertainty in the inference is described by the conditional distribution $p(x_1^{t-1}, x_2^{t-1}|x_1^t, x_2^t)$. In a general nonlinear deterministic system, $p(x_1^{t-1}, x_2^{t-1}|x_1^t, x_2^t)$ is supported by a finite number of points in attractor $A$, except for some special cases (such as that either function $f$ or $g$ is flat). To rephrase, identifying the system's current state $(x_1^t, x_2^t)$ constrains its past state $(x_1^{t-1}, x_2^{t-1})$ on a set of zero-dimensional manifolds (i.e., points). **Figure 1a-ii** depicts a simple case in which the previous state is perfectly identified as a single point.

What if we did not use the joint observation of two nodes' values, $p(x_1^t, x_2^t)$ but rather inferred the past values of individual nodes separately based on marginal observations, $p(x_1^t)$ and $p(x_2^t)$? Because we assumed no uncertainty in observing each node's current value, $p(x_1^t, x_2^t)$, $p(x_1^t)$ and $p(x_2^t)$ are all delta functions, and we thus have $p(x_1^t, x_2^t) = p(x_1^t)p(x_2^t)$. What about the inferred past state, $p(x_1^{t-1}, x_2^{t-1}|x_1^t, x_2^t)$? In fact, such an equality does not hold between the past state distributions, $p(x_1^{t-1}, x_2^{t-1}|x_1^t, x_2^t)$ and $p(x_1^{t-1}|x_1^t)p(x_2^{t-1}|x_2^t)$. Indeed, the previous state generally cannot be identified by the marginal observation when the nodes interact with each other. Namely, $p(x_i^{t-1}|x_i^t)$ is not described by a delta function of $x_i^{t-1}$ ($i = 1,2$), even if $p(x_i^t)$ is a delta function, and thus generally

$$p(x_1^{t-1}, x_2^{t-1}|x_1^t, x_2^t) \neq p(x_1^{t-1}|x_1^t)p(x_2^{t-1}|x_2^t). \tag{3}$$

This fact could be understood intuitively as follows: for example, the set of states that fall in the support of the marginal distribution $p(x_1^t)$ is represented by the vertical red line in **Figure 1a-iv**, reflecting the uncertainty about the current value of $x_2^t$. This set is generally mapped to an oblique line (or a curve) in the previous time point (the red curve in **Figure 1a-iii**) due to the interaction between $x_1$ and $x_2$. Because we do not know where the actual past state was on this curve, we have 1-dimensional uncertainty in the past state of $x_1$ when we consider the projection of this curve onto the $x_1$-axis (as indicated by the non-zero length of the red bar on the horizontal axis in **Figure 1a-ii, iii**). The same argument applies to $x_2$, and thus we have another 1-dimensional uncertainty, now for $x_2$, as shown by the blue bar on the vertical axis. Together, we have a 2-dimensional uncertainty in the past state inferred from the separate ("partitioned") observations $p(x_1^{t-1}|x_1^t)p(x_2^{t-1}|x_2^t)$ in total (as depicted by the dark gray rectangle in **Figure 1a-iii**). Now, recalling that the inference based on a joint observation, $p(x_1^{t-1}, x_2^{t-1}|x_1^t, x_2^t)$, had 0-dimensional uncertainty, we can understand that $p(x_1^{t-1}|x_1^t)p(x_2^{t-1}|x_2^t)$ generally differs from $p(x_1^{t-1}, x_2^{t-1}|x_1^t, x_2^t)$.

This inequality between the joint and partitioned conditional probability distributions is the key for characterizing the integrated information value ($\varphi$) in IIT [1,2,4,14,15]. (Note that basically the same argument applies to the relationships between the current and future states.) How to quantify the difference between the joint and partitioned distributions is arbitrary. Roughly speaking, IIT 2.0 used the Kullback-Leibler divergence [2,15] and IIT 3.0 the earth mover's distance (EMD) [4] to quantify the differences between the joint and partitioned probability distributions. Other information theoretic indices are proposed for practical applications [15–17].

Here, we propose an alternative way of quantifying the difference in distributions based on the topological dimensionality of uncertainty rather than precise information-theoretic quantities. The idea is simple: for example, in the case we described above, $p(x_1^{t-1}, x_2^{t-1}|x_1^t, x_2^t)$ is supported by a 0-dimensional manifold (i.e., point(s)), whereas $p(x_1^{t-1}|x_1^t)p(x_2^{t-1}|x_2^t)$ was supported by a 2-dimensional manifold (i.e., rectangle). Then, we say "the difference between those two distributions is 2 (= 2 – 0)". Formally, if we denote the dimensionality of the support of a distribution $p$ by $\text{Dim}[p]$, the integrated information in terms of the topological dimensionality ($\varphi^{\text{Dim}}$) can be written as follows:

$$\varphi^{\text{Dim}} \equiv \text{Dim}[p(x_1^{t-1}|x_1^t)p(x_2^{t-1}|x_2^t)] - \text{Dim}[p(x_1^{t-1}, x_2^{t-1}|x_1^t, x_2^t)]. \tag{4}$$

In this example,

$$\varphi^{\text{Dim}} = 2. \tag{5}$$

Throughout this paper, we consider cases in which attractors have integer dimensions, and thus $\varphi^{\text{Dim}}$ takes integer values, although we can extend the framework to indexes taking real values by considering non-integer dimensionality such as fractal dimensions [18,19]. Note that, in contrast to the typical frameworks for IIT, the current dimensionality-based argument works in continuous dynamical systems.

## Dimensionality suggests no integration in a mechanistically disconnected system

For a sanity check, let us now consider how the dimensionality-based quantification of the integrated information works in a physically/mechanistically separated system, as shown in **Figure 1b-i**. This system has self-feedback on each node but no interaction between the nodes:

$$x_1^t = f(x_1^{t-1}), \tag{6}$$

$$x_2^t = g(x_2^{t-1}). \tag{7}$$

Although this system could form an apparently 2-dimensional attractor when we plot the trajectory of $(x_1^t, x_2^t)$, it is a product of two smaller dynamical systems. How does this fact affect our index of integrated information $\varphi^{\text{Dim}}$? As before, if we could identify the current state by observing the entire system $(x_1^t, x_2^t)$, the possible past state $(x_1^{t-1}, x_2^{t-1})$ can be constrained to a finite number of points (again, except for some special cases). Then, let us say $\text{Dim}[p(x_1^{t-1}, x_2^{t-1}|x_1^t, x_2^t)] = 0$. How about the partitioned observation, $p(x_1^{t-1}|x_1^t)p(x_2^{t-1}|x_2^t)$? Because there is no interaction between the nodes, each of $x_1$ and $x_2$ forms an autonomous dynamics by itself. This means that observing $x_1$ alone provides sufficient information to constrain the past state to be on 0-dimensional manifold(s), and so does the observation of $x_2$ alone. Namely, the uncertainty in the partitioned observation $p(x_1^{t-1}|x_1^t)p(x_2^{t-1}|x_2^t)$ is still 0-dimensional, and thus the partition does not increase the dimensionality of the uncertainty. This fact is intuitively represented by the red vertical and blue horizontal lines in **Figure 1b-ii** (notice the difference from the case in which there are interactions between the nodes; **Figure 1a-ii**). Therefore, in this system, $\text{Dim}[p(x_1^{t-1}|x_1^t)p(x_2^{t-1}|x_2^t)] = 0$, from which we have

$$\varphi^{\text{Dim}} = 0. \tag{8}$$

This is the desired result. The above example demonstrates that the dimensionality-based index of integrated information, $\varphi^{\text{Dim}}$, correctly captures the "absence of integration" in a disconnected system, just like in the original formulation of IIT.

## Dimensionality suggests integration in an attractor reconstructed from partial observation

We have seen that the dimensionality-based index of integrated information seems to yield reasonable results for simple examples. However, what are the advantages of considering the dimensionality instead of the precise information quantity? To see this, we now turn to considering a case in which we do not observe the entire system but can access only part of it—say, $x_1$ alone (**Figure 1c**). As in the first example, let us assume mutual interaction between the two nodes. In contrast to the previous cases, however, now we assume that we never have access to $x_2$.

Usually, there is no means of measuring the integrated information between $x_1$ and $x_2$ when we cannot observe $x_2$'s state. However, because $x_1$ and $x_2$ are interacting, $x_2$'s information could be implicitly coded by the temporal evolution of $x_1$. If so, we might be able to reconstruct some aspects of the dynamics in the entire system from the observation of its subset. This is indeed the case in nonlinear, deterministic dynamical systems in general—known as the mathematical property called "delay embedding" [11,12]. Delay-embedding theorems claim that, in short, the temporal pattern of a single variable has a smooth one-to-one mapping to the state of the entire system that the observed variable belongs to. In general, if we have an autonomous dynamical system comprising $N$ variables $(x_1, \ldots, x_N)$ that interact with each other, the trajectory of $(x_1^t, \ldots, x_N^t)$ forms an attractor in this $N$-dimensional space. Let us say we can only observe the time series of $x_1^t$. According to the delay-embedding theorems, we can reconstruct the attractor's topology (a shape defined based on connectivity) by plotting the trajectory of $\left(x_1^t, x_1^{t-\tau}, \ldots, x_1^{t-(d-1)\tau}\right)$ instead of $(x_1^t, \ldots, x_N^t)$ when $d$ is sufficiently large, where $\tau$ is called the unit delay and $d$ is called the embedding dimension. It is known that if $d$ is larger than the original attracter's dimensionality, the attractor can be reconstructed almost anywhere on itself with an ignorable volume of overlaps with itself [12]. It might be somewhat surprising that the property of global dynamics could be reconstructed (in a topological sense) solely from the local observation, although it is proven to be the case in almost any type of nonlinear, deterministic dynamical system (and the fact that we can extract the information about the unobserved nodes via embedding may lead to an issue on the definition of "state" in a time series and the meaning of spatial partitioning, as we will reconsider later).

Dimensionality is an aspect of such topological properties reconstructed through the delay embedding. Thus, it is tempting to expect that the present dimensionality-based index of integrated information could be inferred (at least to some extent) from the partial observation. This is indeed possible, although we need a bit of tweaking, as shown below. Now, let us see how this idea works in our simple example (**Figure 1c**).

Because we can only observe $x_1$, we consider a 2-dimensional delay coordinates $(x_1^t, x_1^{t-\tau})$ instead of the original 2-dimensional state space $(x_1^t, x_2^t)$. The unit delay $\tau$ could be chosen arbitrarily. Let $A_1$ denote the reconstructed attractor. Precisely, these 2-dimensinal coordinates are generally not sufficient for embedding when the original attractor $A$'s dimensionality is 2, as the reconstructed attractor $A_1$ includes overlaps with itself. Nonetheless, because the original attractor is 2-dimensional, identifying a 2-dimensional state $(x_1^t, x_1^{t-\tau})$ in these delay coordinates can constrain the original state $(x_1^t, x_2^t)$ within a finite set of points (0-dimensional manifold), as long as the number of self-overlaps is finite (which seems to be the case except in pathological situations). This means that we can also infer the past state in the delay coordinates, $(x_1^{t-1}, x_1^{t-\tau-1})$, with 0-dimensional uncertainty.

As the reader may notice, this situation is quite similar to the case in which we could observe the entire system (**Figure 1a**). Then, what happens if we consider the partitioned observation, just as before? Now, we cannot consider the spatial partition (because we observe only a single node!). Instead, let us introduce a new partition: a "temporal partition." That is, we consider the partition between temporally distant observations $x_1^t$ and $x_1^{t-\tau}$. Applying the same arguments to this temporal partition reveals that observing either of $x_1^t$ and $x_1^{t-\tau}$ alone leads to 1-dimensional uncertainty about the past states $x_1^{t-1}$ and $x_1^{t-\tau-1}$, respectively, and thus the net dimensionality of the uncertainty is 2 (**Figure 1c**). Therefore, the dimensionality-based index of the integrated information in the delay coordinates turns out to be

$$\varphi_1^{\text{Dim}} = 2. \tag{9}$$

Again, the result matches that of the case in which we could observe the entire system (**Figure 1a**). This fact is interesting because it means that we could reach the same result based on two distinct data: one from the complete observation of the entire system and the other from the partial observation of its subset. In particular, both results match the dimensionality of the attractor in the mutually interacting system we considered here.

One may suspect that it is only a coincidence, but in fact, the dimensionality of the reconstructed attractor generally gives an upper bound of $\varphi^{\text{Dim}}$ in the original space. Indeed, when we have a general $d_A$-dimensional attractor formed by $N$ mutually interacting nodes $(x_1, \dots, x_N)$, the attractor can be reconstructed within $d_A$-dimensional delay coordinates of node $x_i$, $\left(x_i^t, \dots, x_i^{t-(d_A-1)\tau}\right)$, allowing $d_A$-dimensional self-overlaps. As long as the number of self-overlaps is finite, the same argument applies, resulting in 0-dimensiosnal uncertainty in inferring the past state with the joint observation. A temporal (bi-)partition, $\left\{\left(x_i^t, \dots, x_i^{t-k+1}\right), \left(x_i^{t-k}, \dots, x_i^{t-(d_A-1)\tau}\right)\right\}$ ($\forall k \in \mathbb{N}$), leads to $d_A - k$ and $k$-dimensional uncertainties for the individual partitioned observations, and thus the net uncertainty turns out to be $d_A$-dimensional. Together, $\varphi_i^{\text{Dim}} = d_A$. On the other hand, $\varphi^{\text{Dim}}$ in the original system is upper-bounded by the attractor dimensions, $d_A$. When $d_A < N$, $\varphi^{\text{Dim}}$ could be smaller than $d_A$. Interestingly, an appropriate projection of the original $N$-dimensional space to a $d_A$-dimensional space (e.g., by clustering the nodes and averaging the node values within each cluster) can recover its upper bound, $\varphi^{\text{Dim}} = d_A$—which is analogous to the fact that integrated information could be maximized by appropriate coarse-graining [20–22].

## Dimensionality suggests the "exclusion" of upstream nodes in an asymmetric interaction

The previous examples show that $\varphi^{\text{Dim}}$ in a mutually interacting system reflects the dimensionality of the attractor (whether the observation is about the whole or partial system), whereas $\varphi^{\text{Dim}} = 0$ in a disconnected system. Note that the apparent dimensionalities of the attractors were both 2 in those examples (**Figures 1a** and **1b**). In this regard, we can interpret $\varphi^{\text{Dim}}$ as an index of "interaction-relevant dimensionality" rather than the apparent dimensionality within the original phase space.

The dimensionality-based characterization becomes even less trivial when we consider a system having a hierarchy in terms of the directionality of interactions. To see this, let us consider an example in which the nodes do not mutually interact but rather have a directed interaction (**Figure 2a–c**). Now, the node $x_2$ affects, but is not affected by, node $x_1$, which can be formally written as

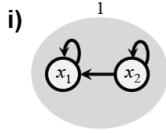
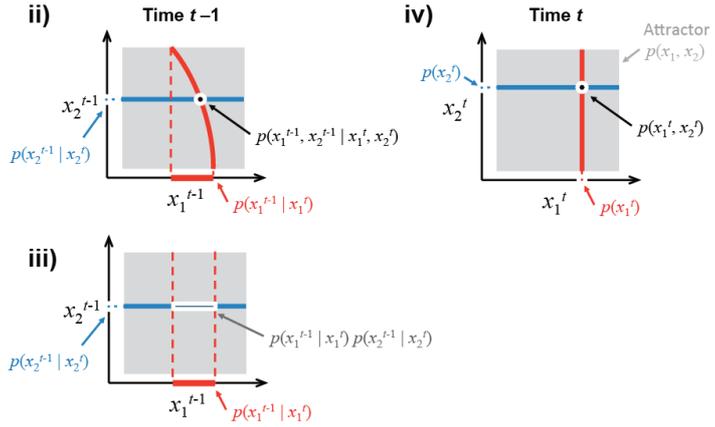
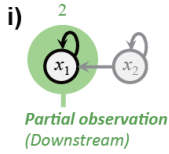
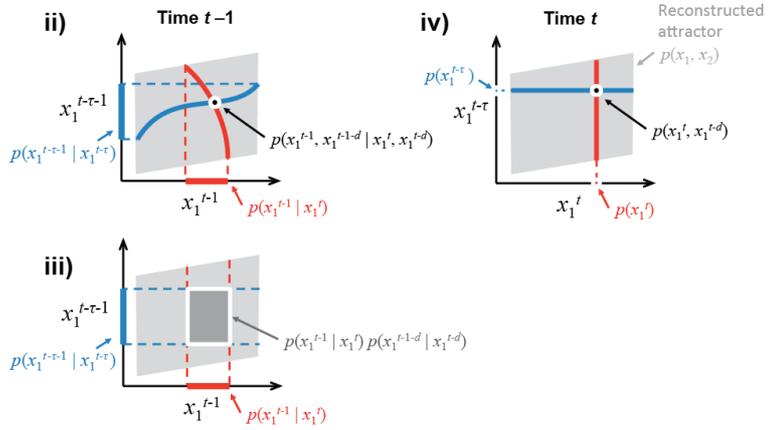
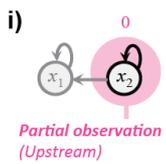
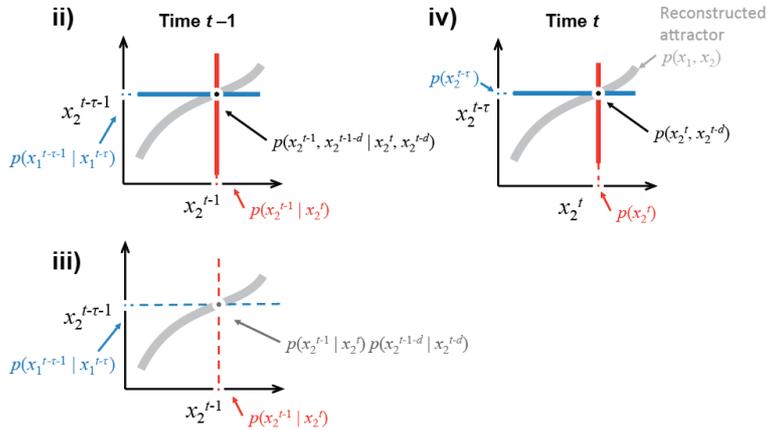

**Figure 2**

Heterogeneity of dimensionality-based index of integrated information under a directed interaction.
(a) The index derived based on the observation of the entire system.
(b) The index derived based on the observation of the downstream node.
(c) The index derived based on the observation of the upstream node.
The inset conventions follow those of **Figure 1**.

$$x_1^t = f(x_1^{t-1}, x_2^{t-1}), \tag{10}$$

$$x_2^t = g(x_2^{t-1}). \tag{11}$$

We can define $x_1$ as the "downstream" and $x_2$ as the "upstream" in the system. Note that the upstream node $x_2$ forms an autonomous dynamical system by itself, whereas the downstream node $x_1$ belongs to the dynamical system formed by both $x_1$ and $x_2$. As we did before, we assume that the system has an apparently 2-dimensional attractor in the phase space of $(x_1, x_2)$.

Let us first consider the simultaneous observation of the entire system (**Figure 2a**). Applying the same analysis as before to this system, we find that identifying the system's current state $(x_1^t, x_2^t)$ constrains its past state $(x_1^{t-1}, x_2^{t-1})$ on a set of zero-dimensional manifolds. On the other hand, in the partitioned observations, identifying the upstream node $x_2^t$ constrains its own past state $x_2^{t-1}$ with 0-dimensinal uncertainty, as it forms a single-node dynamics by itself, whereas identifying the downstream node $x_1^t$ leaves a 1-dimensional uncertainty about its past state $x_1^{t-1}$, reflecting the unknown effect from the upstream. Together, the net uncertainty in the partitioned observation is 1-dimensional, and thus the index of integrated information in the system is

$$\varphi^{\text{Dim}} = 1. \tag{12}$$

Notably, this value of $\varphi^{\text{Dim}}$ under the directed interaction is smaller than that under the mutual interaction (**Figure 1a**), which is qualitatively consistent with the result of the original IIT [2,4].

What if the observation is partial? There are two possibilities of partial observations: observing only the downstream node $x_1$ (**Figure 2b**) or observing only the upstream node $x_2$ (**Figure 2c**). First, when we observe the downstream alone, we can plot the state trajectory in the delay coordinates $(x_1^t, x_1^{t-\tau})$ to reconstruct the topology of the attractor being realized in the entire system $(x_1, x_2)$, which has 2 dimensions in this case (**Figure 2b**). This situation is the same as that in **Figure 1c**, and considering the same temporal partition reveals that the index of integrated information based on this reconstructed attractor is

$$\varphi_1^{\text{Dim}} = 2. \tag{13}$$

On the other hand, when we plot a similar trajectory in the 2-dimensional delay coordinates with the upstream $(x_2^t, x_2^{t-\tau})$, we can reconstruct only a 1-dimensional manifold (**Figure 2c**). Again, it is because the upstream node $x_2$ receives no effect from its downstream, forming a smaller autonomous dynamical system. Because the reconstructed attractor is 1-dimensional, the temporal partition in this 2-dimensional delay coordinates does not increase the uncertainty, resulting in

$$\varphi_2^{\text{Dim}} = 0. \tag{14}$$

To summarize the results presented above, in this system with a directed interaction,

$$\varphi_1^{\text{Dim}} > \varphi^{\text{Dim}} > \varphi_2^{\text{Dim}}. \tag{15}$$

These inequalities illustrate that our dimensionality-based index of integrated information is maximized when it is quantified within the downstream node dynamics, not within the entire system. The integrated information maximized at the downstream agrees, again, with the results of the original formulation of IIT [4]. In particular, the higher integration in a subset of the system rather than the whole demonstrates the axiomatic property of "exclusion" assumed in IIT: namely, the physical substrate of conscious experience has unique borders (e.g., the contents of conscious experience can exclude the phenomenal distinction of feeling one's blood pressure as being high or low) [3–5]. In the present framework, it is natural to interpret the system's subset $i$ that maximizes $\varphi_i^{\text{Dim}}$ as an analogue of 'complex' in IIT, which determines the borders of subjective experiences.

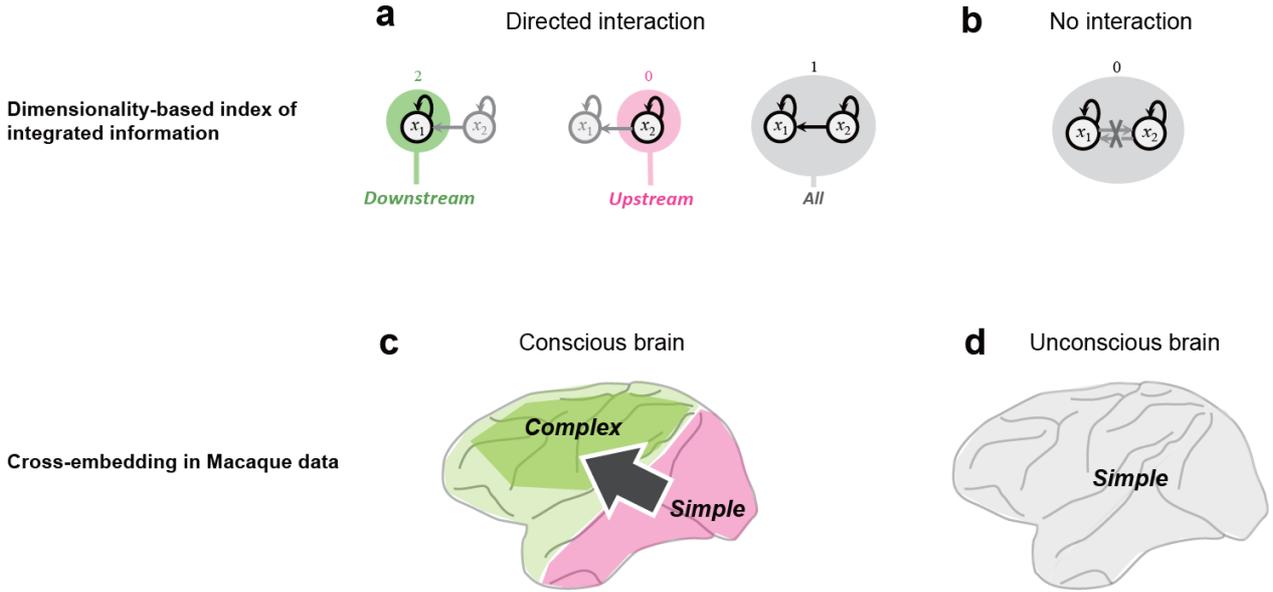

**Figure 3**
Comparison of the dimensionality-based index of integrated information and the interaction-relevant attractor dimensionality ("complexity") revealed by cross-embedding in conscious and unconscious animals.
(a) The system with a directed interaction between nodes (the same as in **Figure 2a–c**).
(b) The system with no interaction (the same as in **Figure 1b**).
(c, d) Summary figures modified from Ref. [13]. (c) The distribution of the attractor complexity revealed by a cross-embedding analysis in awake (conscious) macaque monkeys. (d) The distribution of the attractor complexity revealed by a cross-embedding analysis in anesthetized (unconscious) macaque monkeys.

## Discussion

### Relevance to the cross-embedding complexity

We have seen that our dimensionality-based index of integrated information reflects the dimensions of attractors being realized in dynamical systems in a way sensitive to how the nodes in the systems interact with each other. For example, the value of $\varphi^{\text{Dim}}$ could differentiate a mutual interaction (**Figure 1a**), a directed interaction (**Figure 2a**), and no interaction (**Figure 1b**). An alternative way to quantify the interaction-relevant dimensionality (complexity) of the attractor dynamics is "cross-embedding," which measures the embedding dimensions necessary for inferring a node's value from the temporal pattern of another node's value [13]. Interestingly, analogous to the present dimensionality-based measure of integrated information in the interacting and disconnected systems (**Figures 3a** and **3b**), the cross-embedding indexes higher dimensionality for the downstream nodes than the upstream nodes, both in artificial systems and in the actual brain dynamics in conscious animals, and such strong heterogeneity was not observed in unconscious animals (**Figures 3c** and **3d**) [13]. Indeed, the cross-embedding and the dimensionality-based integrated information share the basic idea that the high-dimensional attractor dynamics realized through interactions are relevant for consciousness. Moreover, similar to the cross-embedding [13], the present index of integrated information demonstrates that information about other nodes can be reconstructed from local dynamics through the delay-embedding technique. This non-localized nature of integrated information can be taken as a form of information 'broadcasting' among nodes, which the Global Neuronal Workspace Theory associates with consciousness [23]. Future studies will investigate more detailed relationships between the cross-embedding and the dimensionality-based integrated information with theoretical analysis and neural recordings.

**Spatial partitions and coordinate transformations**

To assess the dimensionality of dynamics with partial observations, we introduced the idea of "temporal partitioning" in the attractor reconstructed within delay coordinates, based on the delay-embedding theorems. This is a key contribution of this study that could bridge IIT's framework to empirical data, in which we often have access to only partial observations of the studied system. At the same time, the delay embedding and temporal partitioning may invite a new question about the meaning of the spatial partitions considered in the original formulations of IIT: because we can extract information about unobserved variables through delay embedding by regarding the temporal pattern in a subset of the system as a "state," the conclusion derived from the spatial partition could be affected severely by the definition of state within each node. Although it was already partially addressed in a previous study how the spatiotemporal coarse-graining affects the effective information [20], it is yet to be elucidated how the information leveraging by the delay embedding changes the net integrated information. The present dimensionality-based indexing of integrated information allows us to make use of, rather than suffer from, the effects of embedding in continuous dynamical systems. Moreover, the dimensionality-based assessment could be robust to changes in the definition of states because the topological dimensionality is in many cases invariant to coordinate transformations. Note that this invariance is gained in exchange for a more detailed characterization of the information-theoretic quantity; the topological dimensionality is a much coarser measure than the usual measures of integrated information due to the topological invariance, but our point here is that such invariance could allow us to index a form of integrated information even with a partial observation of a system and to bridge that index to the original notions of integrated information.

**Limitations**

Although we believe that the present topology-based approach will provide insights to IIT from practical and theoretical viewpoints, there is still room for elaboration. A major limitation of the current framework is that it assumes that we can estimate the exact dimensionality of the attractors, which can be challenging in real data. To implement the computation described here requires an efficient algorithm for estimating the underlying the attractor dimensionality, but we expect that it should be possible by extending a dimensionality estimation algorithm similar to the one used in the cross-embedding method [13]. Another caveat of the current embedding-based argument is that mathematically rigorous claims can be applied only to continuous, deterministic systems. Although empirical studies with artificial and real data have shown that delay embedding works even in dynamical systems including some stochasticity [13,24–26], future theoretical studies are required for more thorough verifications of the method. Note also that in realistic situations including stochastic dynamics, some information could be lost in the communications among nodes due to noise or other constraints on signal transmissions (e.g., narrow-band temporal frequency responses) that make the downstream information degenerate. In such cases, the attractor dimensions are not always maximized at the system's downstream as in the examples we discussed—which agrees with our intuition that the maximally integrated information should be observed in the central nervous system, rather than its peripheral downstream (e.g., muscles).

Lastly, although the present study focused on the attractor dimensions and relating them to the integrated information quantity as a substrate for the level of consciousness, it remains to be investigated how we can characterize the quality (or contents) of consciousness within this topological framework. A potentially useful approach to characterizing the quality of consciousness is to look at more detailed structures of the attractors, such as the number of holes in each dimension or the higher-order relationships among multiple attractors reconstructed from the individual nodes' dynamics.

**Conclusion**

Currently, the value of IIT is still a subject of debate, attracting both enthusiasm and criticism [27]. An important next step would be to test the fundamental concepts of IIT empirically. For practical and theoretical reasons, however, it has been difficult to perform a rigorous computation of integrated information from real neuronal data. Our present study offers one practical measure of integrated information from real neural data in which the observations are partial and the variables are continuous. Specifically, we have shown that in continuous attractor dynamics, the topological dimensionality of a reconstructed attractor can be used to index the degree of integrated information. We believe that this captures a critical aspect of integrated information as it is invariant to general coordinate transformations. This topological dimensionality-based characterization is not only consistent with the existing framework of IIT, but it also significantly relaxes the conditions required for evaluating the integrated information. As such, the topological dimensionality enables us to assess the integrated information even from partial observations and provides a much-needed framework for testing the theory with experimental data.